\documentclass[
preprint,
nofootinbib,
aps,pra,
]{revtex4-1}
\usepackage{graphicx}
\usepackage{amssymb}
\usepackage{amsmath}
\usepackage[usenames,dvipsnames]{color}
\usepackage{bm}
\usepackage{epstopdf} 
\begin{document}
\title{Polaronic effects of an impurity in a Fermi superfluid away from the BEC limit}
\author{G. Lombardi}
\email{giovanni.lombardi@uantwerpen.be}
\affiliation{TQC, Universiteit Antwerpen, Universiteitsplein 1, B-2610 Antwerpen, Belgium}
\author{J. Tempere}
\affiliation{TQC, Universiteit Antwerpen, Universiteitsplein 1, B-2610 Antwerpen, Belgium}
\affiliation{Lyman Laboratory of Physics, Harvard University, Cambridge, MA 02138, USA}
\date{18/11/2015}
%
%
%
\begin{abstract} 
In this article we study the interaction between an impurity and the gas of Bogoliubov excitations of a Fermi superfluid by mapping it on the polaron problem for an impurity in a BEC. The description of the Fermi superfluid across the BEC-BCS crossover regime is based on a recently developed effective field theory presented in [Eur. Phys. J. B {\bf{88}}, 122 (2015)] and provides us with the interaction-dependent dispersion relations for the Bogoliubov excitations. The behavior of the polaronic coupling constant $\alpha$ and of the effective mass of the polaron is examined in a broad window of the BEC-BCS crossover.
\end{abstract} 

\maketitle 

\section{Introduction}
The polaron concept was first introduced by Landau \citep{THLandau} as a quasiparticle consisting of an electron and the polarization cloud that it drags along while moving in a polar crystal. Since then, many different physical systems -- ranging from solid state to high energy physics -- have been mapped on the polaron problem. Among these realizations, one that has been the focus of much attention in the recent years is the BEC polaron, i.e. a quasiparticle arising from the interaction of an impurity with the Bogoliubov excitations of a Bose-Einstein condensate.
 The theoretical descriptions of the polaron problem can be classified according to the strength of the impurity-boson interaction for which they are valid. The weak coupling regime has been mostly studied by means of a perturbative treatment first developed by Fr\"ohlich \citep{THFrohlich} or by a canonical transformation proposed by Lee-Low-Pines \citep{THLeeLowPines} based on a suggestion by Tomonaga \cite{THAlexandrovDevreese,THDevreese}.
For the strong coupling regime the treatments, introduced by Landau and Pekar \citep{THLandauPekar} and by Bogoliubov and Tyablikov \citep{THBogoliubovTyablikov}, are based respectively on the use of a trial localized wavefunction and again on a canonical transformation.
In addition an all-coupling theory was developed by Feynman based on the path integral formalism \citep{THFeynman}, the results of which were more recently reproduced by using the diagrammatic Monte Carlo method \citep{THProkofevSvistunov}. All these treatments were later applied to the case of the BEC polaron -- see for example \citep{THHuangWan,THNovikovOvchinnikov} for weak coupling, \citep{THLeeGunn,THCucchiettiTimmermans,THKalasBlume,THSachaTimmermans,THRobertsRica} for strong coupling, and \citep{THTCPRB} for all coupling. 
Also a renormalization group all-coupling study \citep{THGrusdtDemler} was proposed. Recently an experimental setup suitable to investigate the BEC polaron was engineered consisting of Cs neutral impurities coupled to a Rb Bose-Einstein condensate \citep{EXPHohmannKindermannWidera}. In the context of Fermi gases, polaronic effects are expected in highly imbalanced Fermi gases in the extreme limit of a single down-spin particle coupled to an ideal gas of up-spins; the so called Fermi polaron has been examined both from a theoretical \citep{THProkofevSvistunovF,THPunkDumitrescu,THMathyParish,THMassignan} and from an experimental point of view \citep{EXPSchirotzekWu,EXPChevyMora}.
In this paper we propose a different version of the polaron problem in a Fermi system. 
In particular we consider the interaction of a single impurity atom with the collective excitations of a fermionic superfluid by mapping it on the same Hamiltonian used in the BEC polaron case. This ansatz is in principle valid only in the extreme BEC side of the Feshbach resonance where the main contribution to the physics of the system should come from the Bogoliubov excitations on top of a molecular BEC.
In the framework of a recently developed effective field theory \citep{THKTLDEpjB} this molecular condensate is described by a macroscopic wavefunction.
The description in terms of a macroscopic wavefunction remains valid also when moving away from the BEC limit and towards unitarity, provided the coefficients of the field equation are properly adapted. 
This allows to calculate the dispersion relation for the Bogoliubov excitations of the superfluid, accounting for the effect of interaction, as the system goes across the BEC-BCS crossover.
In turn, this allows to study how the properties of the BEC polaron change when the underlying condensate no longer consists of pointlike bosons, but of Cooper pairs. The polaron problem is then studied in the weak coupling limit by employing the well known $T=0$ perturbative treatment and the behavior of effective mass and polaronic coupling constant is examined as function of the impurity-boson interaction and of the fermion-fermion interaction in the underlying superfluid. In experiments investigating impurities in Bose Einstein condensates, the polaronic coupling constant can be tuned by acting on the bare boson-boson and boson-impurity scattering lengths. Although methods have been proposed that could boost the polaronic coupling constant and make the strong-coupling regime accessible \citep{EXPHohmannKindermannWidera,EXPShadkhooBruinsma}, up to now only the weak coupling situation has been achieved, hence motivating our focus on this interaction regime.\\
The present article is organized as follows: Section \ref{theory} is devoted to the outline of the theoretical model and the adaptation of the BEC Hamiltonian to the Fermi system. In section \ref{results} the behavior of the polaronic coupling constant $\alpha$ and of the polaron effective mass are analyzed in a wide window of the BEC-BCS crossover. Finally section \ref{conclusions} hosts the conclusions and final remarks.

\section{Theoretical model} \label{theory}
\begin{figure}[h]
\includegraphics[scale=0.7]{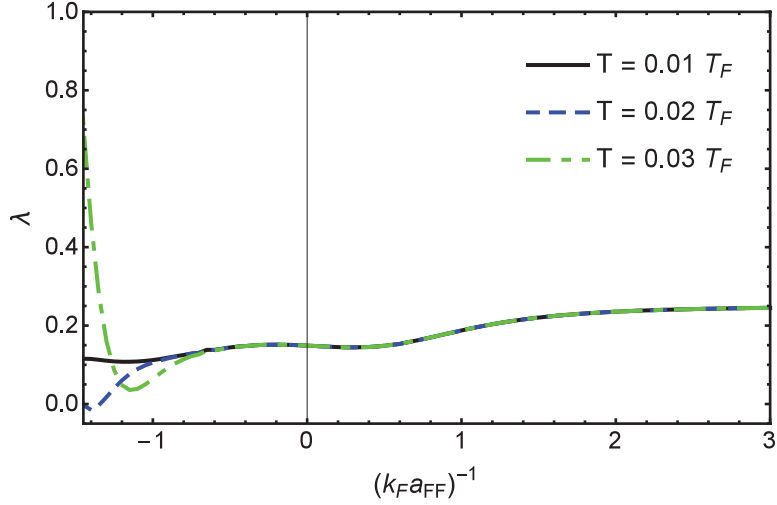}
\caption{Coefficient $\lambda$ as a function of the interaction parameter across the BEC-BCS crossover for different values of the temperature. In the domain of values considered the data for $T=0$ do not differ from those for $T=0.01T_F$}\label{figlambda}
\end{figure}

The problem of a single impurity in a Bose-Einstein condensate can be described by an Hamiltonian of the form
\begin{align}
\hat{H}= E_{GP}+g_{IB}N_0+\frac{\hat{p}^2}{2m_I}+\sum_{\bm{q}}\hbar\omega_{\bm{q}}\hat{\alpha}_{\bm{q}}^\dag\hat{\alpha}_{\bm{q}}+g_{IB}\sqrt{N_0}\sum_{\bm{q}}\sqrt{\frac{\epsilon_{\bm{q}}}{\hbar \omega_{\bm{q}}}}e^{-\mathrm{i}\bm{q}\cdot\hat{\bm{r}}}\left(\hat{\alpha}_{\bm{q}}+\hat{\alpha}_{-\bm{q}}^\dag\right) \label{HIBEC1}
\end{align}
where $E_{GP}$ represents the Gross-Pitaevskii energy of the condensate, $N_0$ is the number of particles in the condensate,
 $\frac{\hat{p}^2}{2m_I}$ is the kinetic energy of the impurity of mass $m_I$, and $\epsilon_{\bm{q}}=\frac{\hbar^2 q^2}{2m_B}$ is the dispersion for a free boson of mass $m_B$. In the last two terms, $\alpha^\dag_{\bm{q}}$ $\left(\alpha_{\bm{q}}\right)$ and $\omega_{\bm{q}}$ are respectively the creation (annihilation) operators and dispersion relation for the Bogoliubov excitations of the bosonic condensate (that play the role of the phonons in analogy with the solid state Fr\"ohlich polaron case). 
The boson-impurity and boson-boson contact interactions are assumed to be $s$-wave and are governed respectively by the coupling constants $g_{IB}$ and $g_{BB}$ that can be related to the corresponding scattering lengths $a_{IB}$ and $a_{BB}$ trough the solution of the Lippmann-Schwinger equation.
\\ In the present treatment the starting point is a fermionic superfluid. In the context of the effective field theory presented in \citep{THKTLDEpjB},  such a system is described in terms of the bosonic superfluid order parameter $\Psi$ by the action
\begin{align}
S(\beta)=&\int_0^\beta\mathrm{d}\tau\int\mathrm{d}\bm{r}\bigg[\frac{\mathcal{D}}{2}\left(\bar{\Psi}\frac{\partial \Psi}{\partial\tau}-\frac{\partial\bar{\Psi}}{\partial \tau}\Psi\right)+\tilde{\mathcal{Q}}\frac{\partial\bar{\Psi}}{\partial\tau}\frac{\partial\Psi}{\partial\tau}-\frac{\mathcal{R}}{2\left|\Psi\right|^2}\left(\frac{\partial \left|\Psi\right|^2}{\partial\tau}\right)^2+\nonumber\\
&\qquad\qquad\qquad+\Omega_s+\frac{\tilde{\mathcal{C}}}{2m_F}\left(\nabla_{\bm{r}}\bar{\Psi}\cdot \nabla_{\bm{r}}\Psi\right)^2 -\frac{\mathcal{E}}{2m_F\left|\Psi\right|^2}\left(\nabla_{\bm{r}}\left|\Psi\right|\right)^2\bigg] \,,\label{Sbeta}
\end{align}
where $\beta$ is the inverse temperature and  $m_F$ is the mass of the fermions. The analytic expressions for the coefficients of the EFT are given in \citep{THKTLDEpjB} in terms of the order parameter $\Psi$, chemical potential $\mu$, imbalance parameter $\zeta$, and fermion-fermion inverse scattering length $1/a_{FF}$. To make this paper self-contained, these expressions are included in the appendix.\\
The Hamiltonian \eqref{HIBEC1} is assumed to remain valid for the description of the fermionic system in the BEC regime i.e. for (large) positive values of $1/a_{FF}$. As mentioned above the goal of this paper is to describe the system away from the BEC limit by employing the Hamiltonian \eqref{HIBEC1} with a modified dispersion relation for the bosonic excitation modes and with a modified condensate density. Both the dispersion relation and the condensate density now depend on the fermionic interaction strength $1/a_{FF}$. The number of particles in the condensate $N_0$ in \eqref{HIBEC1} is calculated via the appropriate expression for a fermionic system that, at saddle point level, reads
\begin{align}
N_0 = V\,n_0 = V\left|\Psi\right|^2\int\frac{\mathrm{d}\bm{k}}{(2\pi)^3}\frac{1}{4E_{\bm{k}}^2}\left(\frac{\sinh\left(\beta E_{\bm{k}}\right)}{\cosh\left(\beta E_{\bm{k}}\right)+\cosh\left(\beta \zeta\right)}\right)^2
\end{align}
where $V$ is the system volume, and the Bogoliubov dispersion $E_{\bm{k}}$ is given by $E_{\bm{k}}=\sqrt{\left(\hbar^2k^2/2m_F-\mu\right)^2+\left|\Psi\right|^2}$.\\
For what concerns the phonon dispersion $\omega_{\bm{q}}$, in \citep{THKTLDEpjB} the spectrum of collective excitations for an ultracold Fermi gas was calculated up to first order in the momentum $\bm{q}$. The same treatment with the inclusion of terms of second order in $\bm{q}$ leads to the dispersion relation
\begin{align}
\hbar\omega_{\bm{q}}=\hbar q\sqrt{v_S^2+\lambda\left(\frac{\hbar q}{2m_F}\right)^2}\label{omegaq0}
\end{align}
With the introduction of the interaction-dependent mass for the bosonic excitation $m_B(\lambda)=m_f/\sqrt{\lambda}$ and the characteristic length $\xi\equiv \frac{\hbar}{\sqrt{2}m_B(\lambda)v_S}$, the last expression becomes
\begin{align}
\hbar\omega_{\bm{q}}=\frac{\hbar^2}{2m_B(\lambda)}q\sqrt{q^2 +2/\xi^2} \label{omegaq}
\end{align}
By determining the quantities $v_S$ and $\lambda$  appearing in \eqref{omegaq0} in terms of the coefficients of the EFT as
\begin{align}
v_S=&\sqrt{\frac{1}{m_F}\frac{\mathcal{U}\tilde{\mathcal{C}}}{\tilde{\mathcal{D}}^2+2\mathcal{U}\tilde{\mathcal{Q}}}}\label{vs}\\
\lambda=&\tilde{\mathcal{D}}^2\frac{\tilde{\mathcal{C}}\left[\mathcal{C}(\tilde{\mathcal{D}}^2+4\mathcal{UR})-2\mathcal{E}(\tilde{\mathcal{D}}^2+4\mathcal{UQ})\right]}{\left(\tilde{\mathcal{D}}^2+2\mathcal{U}\tilde{\mathcal{Q}}\right)^3}.\label{lambda}
\end{align}
we can describe how $\omega_{\bm{q}}$ changes as the system goes across the BEC-BCS crossover.
The quantity $v_S$ can be easily related to the sound velocity. The coefficient $\lambda$, as it becomes clear from the definition of $m_B(\lambda)$, can be instead interpreted as a correction to the mass of the collective excitation. Fig. \ref{figlambda} shows the behavior of this quantity across the BEC-BCS crossover for different temperatures: in the BEC limit the value of $\lambda$ tends to $1/4$ thus making the mass of the bosonic excitation tend to the expected BEC value of $m_B^{(BEC)}=2m_F$. Given this consideration we use the quantity $m_B(\lambda)$ to define the energy $\epsilon_{\bm{q}}$  (that in the BEC polaron case represented the dispersion for a free boson) as $\epsilon_{\bm{q}}\equiv\frac{\hbar^2 q^2}{2m_B(\lambda)}$.\\Finally it has to be remarked that in principle both the boson-boson and impurity-boson scattering lengths $a_{BB}$ and $a_{IB}$ could be related to the fermion-fermion scattering length (see for example \citep{THPetrovSalomonShlyapnikov,THPieriStrinati,THSalasnichBighin}) but this would require a systematic treatment that lies beyond the scope of the present work. However, these quantities will combine into a dimensionless coupling strength, as a function of which we will study the results of our formalism.

\subsection*{Weak coupling limit for an impurity in a BEC condensate}

In order to study the weak coupling regime for the system consisting of an impurity interacting with the collective Bogoliubov excitations of a fermionic superfluid at $T=0$ we employ second order perturbation theory \citep{THHuangWan}. The operator part of the Hamiltonian \eqref{HIBEC1} is divided in an unperturbed part
\begin{align}
\hat{H}_0=\frac{\hat{p}^2}{2m_I}+\sum_{\bm{q}}\hbar\omega_{\bm{q}}\hat{\alpha}_{\bm{q}}^\dag\hat{\alpha}_{\bm{q}}
\end{align} 
accounting for the kinetic energy of the free impurity and the gas of non-interacting collective excitations, plus a perturbation component
\begin{align}
\hat{V}=g_{IB}\sqrt{N_c}\sum_{\bm{q}}\sqrt{\frac{\epsilon_{\bm{q}}}{\hbar \omega_{\bm{q}}}}e^{-\mathrm{i}\bm{q}\cdot\hat{\bm{r}}}\left(\hat{\alpha}_{\bm{q}}+\hat{\alpha}_{-\bm{q}}^\dag\right)
\end{align} 
We start from an unperturbed state of the form $\left |\psi_{\bm{k}}\right\rangle\left|\emptyset\right\rangle$ consisting of a free impurity described by a plane-wave eigenfunction $\psi_{\bm{q}}=e^{\mathrm{i}\bm{k}\cdot\bm{r}}/\sqrt{V}$ and the vacuum state for the phonons $\left|\emptyset\right\rangle$, with energy $E^{(0)}_{\bm{k}}=\left\langle\emptyset\right|\left\langle\psi_{\bm{k}}\right|H_0\left|\psi_{\bm{k}}\right\rangle\left|\emptyset\right\rangle=\frac{\hbar^2k^2}{2m_I}$.
The first order correction to the energy $\Delta E_{\bm{k}}^{(1)}$ is identically zero while the second order correction $\Delta E_{\bm{k}}^{(2)}$ is
\begin{align}
\Delta E_{\bm{k}}^{(2)}&=\sum_{\left|exc\right\rangle \neq \left|\psi_{\bm{k}}\right\rangle\left|\emptyset\right\rangle}\frac{\left|\left\langle exc \right|\hat{V}\left|\psi_{\bm{k}}\right\rangle\left|\emptyset\right\rangle\right|^2}{E_{\bm{k}}^{(0)}-E_{exc}^{(0)}}\, .
\end{align}
The only excited states $\left|exc\right\rangle$ contributing to this quantity are those consisting of the free impurity plus a single phonon, therefore the second order energy correction is
\begin{align}
\Delta E_{\bm{k}}^{(2)}&=N_0 g_{IB}^2\sum_{\bm{q}}\frac{\sqrt{\frac{q^2}{q^2+2/\xi^2}}}{\left[\frac{\hbar^2\bm{k}\cdot\bm{q}}{m_I}-\frac{\hbar^2q^2}{2 m_I}-\frac{\hbar^2}{2m_B}q\sqrt{q^2+\frac{2}{\xi^2}}\right]}\, ,
\end{align}
where in the last line we have introduced the expression for the Bogoliubov dispersion $\omega_{\bm{q}}$ in terms of the healing lenght $\xi$ given in \eqref{omegaq}.
Substituting the sum over momenta $\bm{q}$ with an integral and expanding the integrand in powers of the momentum $k$ of the impurity leads to
\begin{align}
\Delta E_{\bm{k}}^{(2)}&=N_0\left(g_{IB}^{(0)}\right)^2\frac{V}{(2\pi)^2}\int_0^\infty\!\mathrm{d}Q \,2Q^2\times\nonumber\\&\times
\left[-\frac{\sqrt{\frac{Q^2}{Q^2+\frac{2}{\xi^2}}}}{\frac{\hbar^2 Q^2}{2 m_I}+\frac{\hbar^2}{2m_B(\lambda)^2} Q\sqrt{Q^2+\frac{2}{\xi^2}}}-\left(\frac{\hbar^2 Q \,K}{2 m_I}\right)^2\frac{\sqrt{\frac{Q^2}{Q^2+\frac{2}{\xi^2}}}}{3\left(\frac{\hbar^2 Q^2}{2 m_I}+ \frac{\hbar^2}{2m_B(\lambda)} Q\sqrt{Q^2+\frac{2}{\xi^2}}\right)^3}+\cdots\right]
\end{align}
where the dimensionless variables $Q$ (and $K$) are defined as $Q=\xi q$ (and $K=\xi k$).
The term constant in $K$ is divergent for large values of $Q$. This divergence is removed by including the regularized form of the boson-impurity coupling constant $g_{IB}$.\\
The solution of the Lippmann-Schwinger equation up to second perturbative order gives 
\begin{align}
g_{IB}&=\frac{2\pi \hbar^2 a_{IB}}{V m_R(\lambda)}+\alpha\frac{\epsilon_0}{4\pi}\left(\frac{m_I}{m_R(\lambda)}\right)^2\int\mathrm{d}Q\frac{m_R(\lambda)}{m_I} \label{gLP}
\end{align}
where we have introduced, in analogy with the case of an impurity in a BEC, the modified reduced mass $m_R(\lambda)=\left(\frac{1}{m_B(\lambda)}+\frac{1}{m_I}\right)^{-1}$, the energy unit $\epsilon_0=\frac{\hbar^2}{m_I\xi^2}$, and the interaction parameter
\begin{align}
\alpha=\frac{a_{IB}^2}{a^*\xi} \label{alphadef}
\end{align}
The quantity $a^*$ is defined as $a^*=1/\left(16\pi n_c v_S^2/\epsilon_0^2\right)$: in analogy with the BEC polaron case \citep{THTCPRB} -- where the polaronic coupling constant is defined as $\alpha=a_{IB}^2/(a_{BB}\xi)$ --  we expect it to give a measure of the scattering length between the fermion pairs forming the superfluid.  
Substituting \eqref{gLP} and \eqref{alphadef} in the term $N_0 g_{IB}$ of the Hamiltonian provides us with the regularization necessary to have a converging integral for the energy that now reads
\begin{align}
E_{\bm{k}}^{(2)}&=E_{GP}+\frac{2\pi \hbar^2 a_{IB}}{m_R(\lambda)}n_0+\frac{K^2}{2}\epsilon_0+\nonumber\\&+\alpha\frac{\epsilon_0}{4\pi}\left(\frac{m_I}{m_R(\lambda)}\right)^2\int_0^\infty\!\mathrm{d}Q \,Q^2\times\nonumber\\&\times
\left[
\frac{m_R(\lambda)/m_I}{ Q^2}-\frac{\sqrt{\frac{Q^2}{Q^2+2}}}{Q^2+\frac{m_I}{m_B(\lambda)}Q\sqrt{Q^2+2}}-K^2Q^2\frac{\sqrt{\frac{Q^2}{Q^2+2}}}{3\left(Q^2+\frac{m_I}{m_B(\lambda)}Q\sqrt{Q^2+2}\right)^3}+\cdots\right] \label{E2}
\end{align}
It is important to notice that the previous expression is consistent with the theoretical predictions for the weak coupling BEC polaron otained from the all-coupling Feynman treatment: see for reference equation (22) in \citep{THTCPRB}.

\section{Interaction parameter and effective mass of the polaron} \label{results}
\begin{figure}[h]
\includegraphics[scale=0.7]{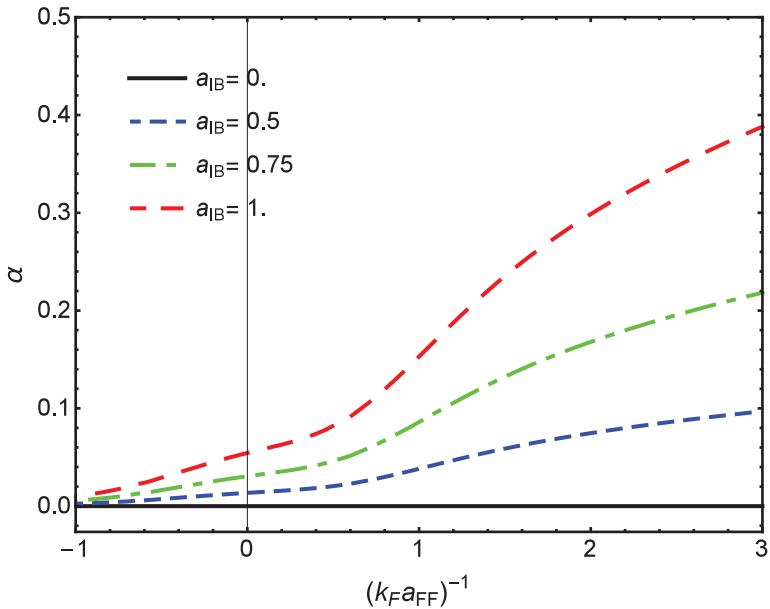}
\caption{Dependence of the dimensionless interaction parameter $\alpha$ on the fermion-fermion interaction strength $(k_Fa_{FF})^{-1}$ across the BEC-BCS crossover for different values of the boson-impurity scattering length at $T=0$}\label{figalphaka}
\end{figure}

\begin{figure}[h]
\includegraphics[scale=0.7]{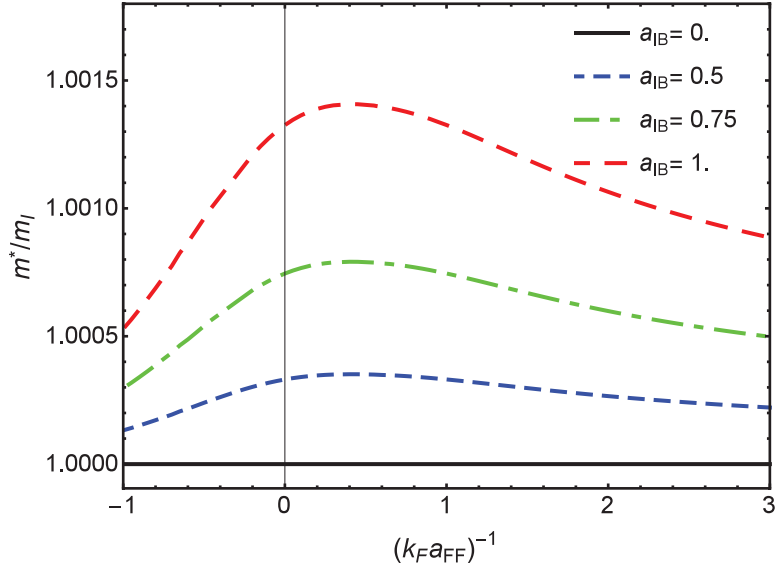}
\caption{Ratio between the effective mass of the polaron and the mass of the impurity as a function of the fermionic interaction parameter $(k_Fa_{FF})^{-1}$ for different values of $a_{IB}$ at $T=0$.}\label{figmeffka}
\end{figure}

As it is clear from \eqref{E2}, the dimensionless parameter $\alpha$ -- often referred to as the polaronic coupling constant -- is the quantity that determines the magnitude of the perturbative corrections to the energy. Figure \ref{figalphaka} depicts its dependence on the fermion-fermion interaction parameter $(k_Fa_{FF})^{-1}$ in the BEC-BCS crossover. A monotonic increase is found for $\alpha$ as the system approaches the BEC side of the Feshbach resonance. Moreover its value at fixed $(k_Fa_{FF})^{-1}$ increases with $a_{IB}$. As espected, when the boson-impurity scattering length $a_{IB}$ is equal to zero $\alpha$ is also identically zero as the impurity does not interact with the superfluid.\\
From the expression for the energy \eqref{E2}, also the effective mass of the polaron $m^*$ can be calculated by using the definition
\begin{align}
\frac{1}{m^*}=\frac{1}{\hbar^2}\left.\frac{\partial^2 \left(E_{\bm{k}}^{(2)}\right)}{\partial k^2}\right|_{k\rightarrow 0}
\end{align}
Inserting the explicit expresion \eqref{E2} for $E_{\bm{k}}^{(2)}$ in the last equation and solving for $m^*$ we obtain
\begin{align}
m^*=m_I\left(1- \alpha\frac{\epsilon_0}{4\pi}\left(\frac{m_I}{m_R(\lambda)}\right)^2\int_0^\infty\!\mathrm{d}Q \,
Q^4\frac{\sqrt{\frac{Q^2}{Q^2+2}}}{3\left(Q^2+\frac{m_I}{m_B(\lambda)}Q\sqrt{Q^2+2}\right)^3}\right)^{-1}
\end{align}
Figure \ref{figmeffka} shows the behavior of the ratio between the effective mass of the polaron and the mass of the impurity across the BEC-BCS crossover for fixed values of $a_{IB}$. A maximum for the ratio $m^*/m_I$ is found for small positive values of the interaction parameter $(k_F a_{FF} )^{-1}$. Similar to the case of the interaction parameter $\alpha$, as could be intuitively expected, also the value of the effective mass at fixed fermion-fermion interaction strength increases with the boson-impurity scattering length.\\
From both Fig.\ref{figalphaka} and Fig.\ref{figmeffka} it appears that a region of major relevance in the domain of values of the interaction parameter is the one around $(k_Fa_{FF})^{-1}\sim 0.4$. For the polaronic coupling constant $\alpha$ this is the region where a marked change in the slope of the curves in Fig.\ref{figalphaka} is observed. On the other hand, considering the behavior of the effective mass, from Fig.\ref{figmeffka} we notice that the maximum of the ratio $m^*/m_I$ is also localized around this position. The importance of this region of the interaction parameter domain was already pointed out in \citep{THLvAKT} where a peak in the inverse pair coherence length was detected suggesting a direct link between the appearence of particular features in this range of values of $(k_Fa_{FF})^{-1}$ and the intrinsic nature of the system.
\section{Discussion and conclusions} \label{conclusions}
In this paper we have studied a system composed by a single impurity atom interacting with the collective excitations of a fermionic superfluid by employing  the Fr\"ohlich-like Hamiltonian widely used to study the similar BEC polaron problem and extending its validity -- in principle limited to the extreme BEC side of the Feschbach resonance -- to a wider region of the BEC-BCS crossover. 
This was done by calculating the $(k_F a_{FF})^{-1}$-dependent form of the dispersion relations for the Bogoliubov excitations of the Fermi superfluid in the context of a recently developed effective field theory \citep{THKTLDEpjB}. 
The system was studied in the weak coupling regime in perturbation theory. In order for this kind of treatment to be valid we had to restrict the analysis to the $T=0$ situation. However, as discussed in \citep{THLvAKT} in regard to dark solitons, this requirement on the temperature introduces a limitation on the reliability of the EFT away from the BEC limit. 
Given this consideration we remark that the results at unitarity and in the BCS regime can be seen just as qualitative predictions.\\ The main focus of this paper was the calculation of the effective mass of the polaron and the analysis of its behavior across the BEC-BCS crossover. For a fixed value of the fermion-fermion interaction strength $(k_Fa_{FF})^{-1}$ the effective mass is shown to increase with the impurity-boson scattering length. The behavior across the BEC-BCS crossover is not monotonic: in particular a broad peak in the value of $m^*$ is found on the near BEC side of the resonance. In the extreme BEC limit the results of the BEC polaron problem are correctly retreived. The polaron effective mass has already been succesfully measured in experiments on ionic crystals and polar semiconductors \citep{EXPLittonButton}. The recent proposal of an experimental setup for the study of the BEC polaron \citep{EXPHohmannKindermannWidera} opens the door to the possibilty of measuring this property also in systems like the one considered in the present paper.\\
Also the variation of the polaronic coupling parameter $\alpha$ was studied as a function of the fermion-fermion interaction, finding a monotonic increase as the system goes from the BCS towards the BEC regime.
\acknowledgements We gratefully acknowledge useful discussions with W. Casteels, S. Klimin, J.P.A. Devreese, W. Van Alphen, and N. Verhelst.
This research was supported by the Flemish
Research Foundation (FWO-Vl), project
No. G.0115.12N, No. G.0119.12N, No. G.0122.12N, No. G.0429.15N, No. G0G6616N,
by the Scientific Research Network of the Research
Foundation-Flanders, WO.033.09N, and by the
Research Fund of the University of Antwerp.

\section*{Appendix A: coefficients of the EFT}
\def\theequation{A.\arabic{equation}}
\setcounter{equation}{0}
In this section we give the an overview of the coefficients appearing in the effective field action \eqref{Sbeta}.
It is convenient to write these coefficients in terms of the functions $f_s(\beta,\epsilon,\zeta)$, which are defined as the solutions of
\begin{align}
f_{s}(\beta,\epsilon,\zeta)=\frac{1}{\beta}\sum_{n}\frac{1}{\left[\left(\omega_n-i\zeta\right)^2+\epsilon^2\right]^s}
\end{align}
where $\omega_n$ are fermionic Matsubara frequencies of the form $\omega_n=(2n+1)\pi/\beta$.
The explicit expression for the first of these functions $f_1(\beta,\epsilon,\zeta)$ is given by
\begin{align}
f_1(\beta,\epsilon,\zeta)=\frac{1}{2\epsilon}\frac{\sinh(\beta\epsilon)}{\cosh(\beta\epsilon)+\cosh(\beta\zeta)}
\end{align}
From this, the other $f_s(\beta,\epsilon,\zeta)$ with $s=2,3,...$ can be calculated by using the simple recursion relation
\begin{align}
f_{s+1}(\beta,\epsilon,\zeta)=-\frac{1}{2s\,\epsilon}\frac{\partial f_s(\beta,\epsilon,\zeta)}{\partial \epsilon}
\end{align}
The complete expressions for the coefficients appearing in $S(\beta)$ \eqref{Sbeta} are hence given by
\begin{align}
\mathcal{\tilde{C}}  &  =\int\frac{d\mathbf{k}}{\left(  2\pi\right)  ^{3}%
}\frac{k^{2}}{3m}f_{2}\left(  \beta,E_{\mathbf{k}},\zeta\right)  ,\label{c}\\
\mathcal{D}  &  =\int\frac{d\mathbf{k}}{\left(  2\pi\right)  ^{3}}\frac
{\xi_{\mathbf{k}}}{w}\left[  f_{1}\left(  \beta,\xi_{\mathbf{k}},\zeta\right)
-f_{1}\left(  \beta,E_{\mathbf{k}},\zeta\right)  \right]  ,\label{d}\\
\mathcal{E}  &  =2w\int\frac{d\mathbf{k}}{\left(  2\pi\right)  ^{3}}%
\frac{k^{2}}{3m}\xi_{\mathbf{k}}^{2}~f_{4}\left(  \beta,E_{\mathbf{k}}%
,\zeta\right)  ,\label{ee}\\
\mathcal{\tilde{Q}}  &  =\frac{1}{2w}\int\frac{d\mathbf{k}}{\left(
2\pi\right)  ^{3}}\left[  f_{1}\left(  \beta,E_{\mathbf{k}},\zeta\right)
\right. \nonumber\\
&  \left.  -\left(  E_{\mathbf{k}}^{2}+\xi_{\mathbf{k}}^{2}\right)
f_{2}\left(  \beta,E_{\mathbf{k}},\zeta\right)  \right]  ,\label{qq}\\
\mathcal{R}  &  =\int\frac{d\mathbf{k}}{\left(  2\pi\right)  ^{3}}\left[
\frac{f_{1}\left(  \beta,E_{\mathbf{k}},\zeta\right)  +\left(
E_{\mathbf{k}}^{2}-3\xi_{\mathbf{k}}^{2}\right)  f_{2}\left(  \beta
,E_{\mathbf{k}},\zeta\right)  }{3w}\right. \nonumber\\
&  \left.  +\frac{4\left(  \xi_{\mathbf{k}}^{2}-2E_{\mathbf{k}}^{2}\right)
}{3}f_{3}\left(  \beta,E_{\mathbf{k}},\zeta\right)  +2E_{\mathbf{k}}%
^{2}wf_{4}\left(  \beta,E_{\mathbf{k}},\zeta\right)  \right]  . \label{rr}%
\end{align}
In addition the thermodynamic potential $\Omega_s$ is
\begin{align}
\Omega_{s}\left(  w\right)   &  =-\int\frac{d\mathbf{k}}{\left(  2\pi\right)
^{3}}\left(  \frac{1}{\beta}\ln\left(  2\cosh\beta E_{\mathbf{k}}+2\cosh
\beta\zeta\right)  \right. \nonumber\\
&  \left.  -\xi_{\mathbf{k}}-\frac{w}{2k^{2}}\right)  -\frac{w}{8\pi a_{s}},
\end{align}
and the coefficients $\mathcal{\tilde{D}}$ and $\mathcal{U}$ appearing in \eqref{vs} and \eqref{lambda} are defined as
\begin{align}
\mathcal{U}\left(  w\right)   &  =w\frac{\partial^{2}\Omega_{s}\left(
w\right)  }{\partial w^{2}},\quad\mathcal{\tilde{D}}\left(  w\right)
=\frac{\partial\left[  w\mathcal{D}\left(  w\right)  \right]  }{\partial
w}.
\end{align}

\bibliography{Refs_polaron_theory}


\end{document}